\documentclass[
    ,final            
  ]
  {aipproc}

\layoutstyle{6x9}

\def\fig#1{fig.~{\ref{#1}}}

\def\eqn#1{eq.~(\ref{#1})}

\def\Li{\mathop{\rm Li}\nolimits}
\def\to{\rightarrow}
\def\e{\epsilon}
\def\del{\partial}
\def\nn{\nonumber}
\def\tr{{\rm tr}}
\def\tlambda{{\tilde\lambda}}
\def\ve{\varepsilon}
\def\spab#1.#2.#3{\sandmm#1.#2.#3}
\def\spba#1.#2.#3{\sandpp#1.#2.#3}
\def\spaa#1.#2.#3.#4{\sandmp#1.{#2#3}.#4}
\def\spbb#1.#2.#3.#4{\sandpm#1.{#2#3}.#4}
\def\spa#1.#2{\left\langle#1\,#2\right\rangle}
\def\spb#1.#2{\left[#1\,#2\right]}
\def\spash#1.#2{\vphantom{\hat K}\spa{\smash{#1}}.{\smash{#2}}}
\def\spbsh#1.#2{\vphantom{\hat K}\spb{\smash{#1}}.{\smash{#2}}}
\def\lor#1.#2{\left(#1\,#2\right)}
\newbox\SlashedBox
\def\slashed#1{\setbox\SlashedBox=\hbox{#1}
\hbox to 0pt{\hbox to 1\wd\SlashedBox{\hfil/\hfil}\hss}#1}
\def\hboxtosizeof#1#2{\setbox\SlashedBox=\hbox{#1}
\hbox to 1\wd\SlashedBox{#2}}

\newbox\charbox
\newbox\slabox
\def\s#1{{      
        \setbox\charbox=\hbox{$#1$}
        \setbox\slabox=\hbox{$/$}
        \dimen\charbox=\ht\slabox
        \advance\dimen\charbox by -\dp\slabox
        \advance\dimen\charbox by -\ht\charbox
        \advance\dimen\charbox by \dp\charbox
        \divide\dimen\charbox by 2
        \raise-\dimen\charbox\hbox to \wd\charbox{\hss/\hss}
        \llap{$#1$}
}}
\def\ksl{\s{k}}
\def\Ksl{\s{K}}
\def\lr{\leftrightarrow}

\begin{document}

\noindent
SLAC--PUB--11316
\hfill hep-ph/0507064

\title{Recent Developments in Perturbative QCD
\footnote{Talk presented at the 13th International Workshop on 
Deep Inelastic Scattering (DIS 05), Madison, Wisconsin, April, 2005. 
Research supported by the US Department of Energy under contract
DE--AC02--76SF00515.}
}


\author{Lance J. Dixon}{
  address={Stanford Linear Accelerator Center,
              Stanford University,
             Stanford, CA 94309, USA}
}

\begin{abstract}
I review recent progress in perturbative QCD on two fronts:
extending next-to-next-to-leading order QCD corrections
to a broader range of collider processes, and applying
twistor-space methods (and related spinoffs) to computations
of multi-parton scattering amplitudes.
\end{abstract}

\maketitle


\centerline{ {\it A method is more important than a discovery,} }
\centerline{ {\it since the right method will lead to new and even
more important discoveries.} }
\centerline{ {\it -- L.~D. Landau} }

\section{Introduction}

Asymptotic freedom~\cite{GWP}, for which Gross, Politzer and Wilczek
received the 2004 Nobel Prize in Physics, provides the conceptual
framework for applying perturbative QCD to short-distance-dominated problems
in hadronic physics, such as the deep-inelastic (DIS) scattering
process, the focus of this series of workshops.
Supplemented with the notion of factorization~\cite{Factorization},
and the experimental determination of parton distributions,
perturbative QCD has become the basis for all quantitative theoretical
predictions for large-transverse momentum processes in hadron-hadron
and $ep$ collisions, as well as jet production in $e^+e^-$ annihilation.

One might have thought that by now, with the aid of computers,
perturbative QCD should have been ``reduced to quadratures'',
that is, to a simple exercise in tabulating and numerically evaluating
Feynman diagrams.
Yet it is often the case that the experimental precision exceeds
the theoretical uncertainties, due to unknown higher-order terms
in the perturbation series.
Here I will cover two topics in computational perturbative QCD,
for which there has been a great deal of progress,
although much still remains to be done.

The first topic concerns next-to-next-to-leading order (NNLO)
corrections to collider processes, also the subject of another talk
at this workshop~\cite{Klasen}.  NNLO computations have been available
for a limited number of collider observables for many years,
but only now are the prospects becoming good for extending them
to a broader range of important precision processes at hadron colliders.

The second topic is a rapidly developing one, in which insights gleaned from
the topological string in twistor space proposed by
Witten~\cite{WittenTopologicalString}, and further developments, promise to
provide efficient means for computing tree-level and one-loop QCD amplitudes
with a large number of external partons, as well as vector and Higgs bosons.
These amplitudes are needed for next-to-leading order corrections to a variety
of processes.


\section{Progress at NNLO}
\label{nnlosection}

For most observables, the QCD perturbation series is a slowly converging one.
(Technically it is an asymptotic series, but rarely are there enough terms
available in the series for the distinction to matter quantitatively.) Typical
next-to-leading order (NLO) corrections for collider processes range from 20\%
to 100\%.  Clearly any kind of precision measurement, say at the few percent
level, will require the NNLO terms in the series as well. Examples where this
precision is desirable include the determination of
\begin{itemize}
\item $\alpha_s$ via jet production and event shapes in $e^+e^-$ annihilation
(as well as in $ep$ collisions)
\item parton distributions via DIS, Drell-Yan
production, and high-$p_T$ jet production at hadron colliders
\item electroweak
parameters, such as $M_W$, via $W$ and $Z$ production at hadron colliders
\item
the ``partonic luminosity'' at the LHC~\cite{DPZ}
\item Higgs couplings.
\end{itemize}

The progress of NNLO computations for collider processes can be
charted in terms of the number of physical scales present in the parton-level
cross sections.  The more scales, the more difficult the computation,
but the more flexible the applications.
In perturbative QCD with massless quarks, all relevant
scales are associated with the external kinematics.
(The dependence of the partonic cross sections on the renormalization
and factorization scales can be determined with relatively little
effort, so it can be neglected in this counting.)
Also, dimensional analysis can be used to remove an overall
dimensionful scale from the problem, leaving just the number of
dimensionless ratios.

\subsection{No-scale problems}

\begin{figure}
  \includegraphics[height=.15\textheight]{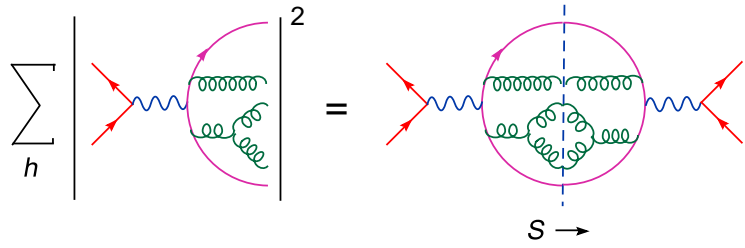}
  \caption{Unitarity relates the $e^+e^- \to \,{\rm hadrons}$
  total cross section to propagator-type loop integrals.}
 \label{UnitaryPropFigure}
\end{figure}

For example, in the total cross section
for $e^+e^-$ annihilation into hadrons, or equivalently the ratio
$R_{e^+e^-}(s)
= \sigma(e^+e^- \to \,{\rm hadrons})
 /\sigma(e^+e^- \to \mu^+\mu^-)$,
the only physical scale is $s$, the square of the center-of-mass
energy.  This scale can be removed trivially, so $R_{e^+e^-}(s)$
is really a no-scale problem, from the computational point of view.
That is, each term in the perturbative series for $R_{e^+e^-}$
is (for fixed renormalization scale $\mu$) a pure number.
Related to the lack of other scales in the
problem is the totally inclusive nature of the observable;
that is, it sums over all hadronic final states with no constraints.
This sum can be performed using unitarity, or the optical theorem,
as illustrated in \fig{UnitaryPropFigure}, transforming the problem
into the computation of the imaginary part of the virtual photon propagator,
or two-point function.

No-scale processes were the first to be computed at NNLO
in perturbative QCD, in the early 1990s.
Besides $R_{e^+e^-}$ and the closely related problem of
the semi-hadronic width of the $\tau$ lepton,
$\Gamma(\tau \to \nu_\tau + {\rm hadrons})$~\cite{Reetau},
various DIS sum rules were also evaluated
at this order: the Bjorken sum rule for neutrino scattering,
$\int_0^1 dx \, [ F_1^{\bar\nu p}(x,Q^2) - F_1^{\nu p}(x,Q^2)]$~\cite{Bjnu},
the Bjorken sum rule for polarized electroproduction,
and the Gross-Llewellyn Smith sum rule for neutrino scattering~\cite{BjGLS}.
The integrals over $x$ not only remove dependence of the observables
on parton distribution functions, but they reduce the computation
to a no-scale, propagator-type problem very similar to $R_{e^+e^-}$.

The technology underlying all these computations was
integration by parts (IBP)~\cite{IBP}.  Total derivatives
of multi-loop integrals in $D=4-2\e$ space-time dimensions
({\it i.e.}, using dimensional regularization),
such as
\begin{eqnarray}
0 &=& \int d^Dp d^Dq \ldots { \del \over \del q^\mu }
 { k^\mu \over p^2 q^2 (p+q)^2 \ldots } \nn\\
  &=& \int d^Dp d^Dq \ldots
 \biggl[ - { 2 k\cdot q \over p^2 [q^2]^2 (p+q)^2 \ldots }
        -  { 2 k\cdot (p+q) \over p^2 q^2 [(p+q)^2]^2 \ldots } \biggr] \,,
\label{IBPeq}
\end{eqnarray}
where $p$ and $q$ are loop momenta, and $k$ is an external
momenta, can be re-expressed as linear equations relating
loop integrals with propagators raised to different powers.
The absorptive parts of four-loop propagators can be related
to three-loop propagators via the $R^*$ operation~\cite{Rstar}.
For the three-loop propagator problem, a recursive solution
to the system of linear IBP equations, in terms of a small set of
irreducible, ``master'' integrals, was implemented in the program
{\tt MINCER}~\cite{MINCER}, making possible
the previously-mentioned NNLO results.

The pure numbers encountered at NNLO in no-scale problems have a very simple
analytic structure. The only algebraic quantities appearing, besides rational
numbers, are the Riemann zeta values, $\zeta(n)$, for $n \leq 5$. The results
from the early 1990s stood as the computational state-of-the-art for many
years. (Very recently, a method for handling the four-loop propagators with all
possible topologies has proven successful~\cite{BCK}, suggesting that the
N$^3$LO results for $R_{e^+e^-}$ may be available before long.)  They have led
to $\alpha_s(Q^2)$ determinations with the smallest theoretical uncertainty ---
if $Q^2$ can be made large enough experimentally, for example, at the $Z^0$
pole. However, the experimental precision is highly stressed by the leading
``1'' in $R_{e^+e^-} \propto 1 + \alpha_s/\pi + \cdots$: A 3\% measurement of
$\alpha_s(M_Z) \approx 0.120$ requires a parts per mil measurement at the $Z$
pole of $\Gamma(Z\to\,{\rm hadrons})/\Gamma(Z \to \mu^+\mu^-)$~\cite{RPP04}.
Observables beginning at order $\alpha_s$, {\it e.g.}, $e^+e^-$ event-shape
variables, are less demanding in this way, motivating their NNLO computation,
which is a multi-scale problem, only now approaching completion.

\subsection{One-scale problems}
\label{onescale}

Also around 1990, the first NNLO computation of a one-scale collider process
was carried out, the total cross section for inclusive production in hadronic
collisions of a lepton pair via the Drell-Yan process, {\it i.e.} via a vector
boson $V = \gamma^*$, $W$ or $Z$~\cite{HvNM}. At the parton level, the process
$pp \to V + X$, where $V$ has mass $M_V$, introduces the single dimensionless
ratio $z \equiv M_V^2/\hat{s}$, where $\hat{s}$ is the squared partonic
center-of-mass energy.  Whereas the NLO correction to the total cross section
was sizable, at NNLO, for $W$ or $Z$ production at the Tevatron or LHC, the
perturbative series nicely stabilized.

The NNLO Drell-Yan result was followed quickly by the Wilson
coefficient functions $C_i(z)$ for DIS structure functions~\cite{NNLODIS}
--- except for the longitudinal structure function $F_L$,
which begins at one order higher in $\alpha_s$, and whose computation
was just completed this spring~\cite{MVVCoeffs}.

In the past few years, the NNLO corrections to two additional one-scale
collider processes were attacked.  First, the total cross section for inclusive
production of a Higgs boson in hadronic collisions, $pp \to H + X$ was computed
in the large $m_t$ approximation. In this limit, Higgs production is
kinematically very similar to the Drell-Yan process, because $V$ and $H$ are
both massive color-singlet particles, and no other mass scales remain in the
problem --- other than the overall Higgs coupling strength, dictated by the
operator $C(m_t) H \tr(G_{\mu\nu}G^{\mu\nu})$. Indeed the first Higgs
computation, via a high-order expansion in $1-z$ (where now $z =
M_H^2/\hat{s}$)~\cite{HarlanderKilgore} was also applied to the Drell-Yan case,
revealing a numerically small correction to the original results.

\begin{figure}
  \includegraphics[height=.15\textheight]{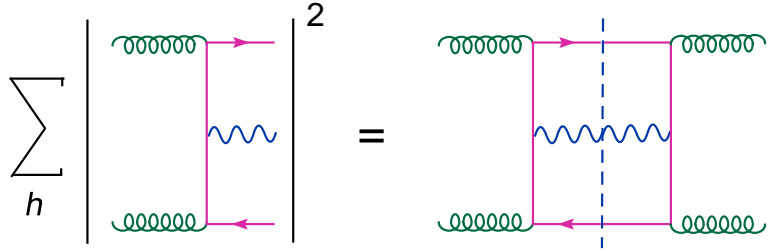}
  \caption{Unitarity relates the $gg \to V + X$ cross section
   to forward-scattering loop integrals.}
 \label{VUnitarityFigure}
\end{figure}

A second Higgs production computation~\cite{AnastasiouMelnikov}
exploited unitarity to express the partonic Higgs cross section
as a forward scattering process, as shown in \fig{VUnitarityFigure}.
In this case, the state that scatters forward is not a single massive
virtual photon, but a pair of massless initial partons.  Also,
not every cut is considered, but only those that cut through the
Higgs particle (or vector boson $V$).   The advantage of this
approach is that the large number of phase-space integrals that
have to be performed (only one example of which is shown on the
left-hand side of \fig{VUnitarityFigure}) can be traded for
multi-loop integrals, to which the IBP method can be applied
in an automated fashion~\cite{LaportaAIR}, in order to
reduce the integrals to a manageable set of master integrals.
In contrast to the no-scale examples, now all the master integrals
depend on $z$.  They also depend on the dimensional regularization
parameter $\e$, and have to be expanded in a Laurent expansion around
$\e=0$, beginning at order $1/\e^3$, due to infrared divergences
in the integrals.  Fortunately, the IBP method also provides a way to determine
the $z$-dependence of each coefficient in the Laurent expansion:
Taking a derivative with respect to $z$ produces an integral
which can also be reduced to master integrals, thus generating a
coupled set of differential equations~\cite{MasterDiffEqs} which
are readily solved in terms of special functions.
In this way the exact dependence of the NNLO partonic Higgs cross
section on $z$ was determined~\cite{AnastasiouMelnikov}
(see also ref.~\cite{RSvN}).

For the Drell-Yan or Higgs total cross section, the special
functions that appear are polylogarithms of the form ${\rm Li}_n(z)$,
defined by $\Li_1(z) = - \ln(1-z)$, and for $n>1$, 
\begin{equation}
\Li_n(z) = \sum_{j=1}^\infty {z^j \over j^n} =
\int_0^z {dt\over t} \Li_{n-1}(t) \,.
\label{PolyLogDef}
\end{equation}
In the Drell-Yan/Higgs application, one needs $n \leq3$; 
the argument $z$ may be replaced by a few other
rational functions of $z$.  The existence of the single scale $z$
allows the analytical complexity, in terms of the number of
different algebraic objects present, to grow significantly with
respect to $R_{e^+e^-}$, but it is not yet out of hand.

In principle, no NNLO computation of a collider process is complete without an
evolution of the parton distributions from low to high scales at the same NNLO
accuracy. Last year saw the long-awaited completion of the NNLO corrections to
the DGLAP evolution kernels for parton distribution functions~\cite{MVVDGLAP},
$P_{ij}(x)$.  These were also computed by considering a forward scattering
problem, with a virtual photon and a quark in the initial and final state. (For
the gluon evolution kernel $P_{gg}(x)$, and also for $P_{qg}(x)$, a fictitious
Higgs-like scalar $\phi$, coupling to gluons via $\phi
\tr(G_{\mu\nu}G^{\mu\nu})$, is used instead of a virtual photon.) After
renormalization and subtraction of collinear divergences from lower loop
orders, the NNLO evolution kernels are identified from the remaining $1/\e$
poles in the expression, which must be subtracted in the $\overline{\rm MS}$
scheme for defining parton distribution functions (or equivalently, leading
twist operators). As a by-product of the computation, the finite, order $\e^0$
terms give the N$^3$LO contributions to the DIS structure function $F_2$ and
the NNLO contribution to $F_L$~\cite{MVVCoeffs}.

For these results, somewhat more complicated special functions are required,
such as $\Li_n(x)$ with $n=4$ for $P_{ij}(x)$ and $n=5$ for $F_2$ and $F_L$.
However, the ordinary polylogarithms are not sufficient; a suitable
generalization, harmonic polylogarithms~\cite{HPL}, can be used instead.  In
fact, these computations were {\it not} performed in terms of the variable $x$,
but rather the variable $N$ appearing in the Mellin transform, $\tilde{f}(N)
\equiv \int_0^1 dx\ x^{N-1}\ f(x)$.
In $N$-space, the special functions that appear at NNLO are harmonic sums, such
as the one-dimensional sum $S_k(N) = \sum_{i=1}^N i^{-k}$, and the
multi-dimensional generalization of it,
\begin{equation}
S_{m,m_1,m_2,\ldots,m_k}(N)
= \sum_{i=1}^N { S_{m_1,m_2,\ldots,m_k}(i) \over i^m } \,.
\label{HarmSum}
\end{equation}
Although it might appear that in Mellin-moment space a no-scale problem has
been recovered, it is of course an infinite set of no-scale problems.  Indeed,
at fixed moment $N$, the program {\tt MINCER} can be used to compute anomalous
dimensions at NNLO for $N=2,4,6,8,10$~\cite{LVMINCER} and even up to
$N=16$~\cite{BVMINCER}. However, for the case of arbitrary $N$, new integral
reduction algorithms had to be developed~\cite{MVVDGLAP}.

Recently, the Mellin moments of the NNLO Drell-Yan and Higgs production
cross sections were presented in terms of harmonic sums of the
form~(\ref{HarmSum})~\cite{BR}, putting the mathematical structure
of the two types of NNLO one-scale problems discussed here on
the same footing.

There have been important phenomenological applications
of these one-scale results.  Also taking into account the
resummation of large threshold logarithms from multiple soft-gluon
resummation, through next-to-next-to-leading logarithmic
accuracy~\cite{CdFGN}, the uncertainty on the total
Higgs production cross section at the Tevatron and LHC has
been considerably reduced, from perhaps 30--40\% at NLO,
to perhaps 10--20\% at NNLO+NNLL.
Some of the immediate phenomenological impact of the NNLO
evolution kernels and coefficient functions on ``DIS-driven''
MRST parton fits~\cite{MRSTDISdriven}
was discussed elsewhere at this meeting~\cite{Klasen,Thorne}.

\subsection{Two or more scale problems}
\label{twoscale}

Although the total cross sections for inclusive Drell-Yan and
Higgs production are now known relatively well theoretically,
in practice such quantities are not measurable experimentally.
Numerous experimental cuts must be imposed to extract a
signal from the background, for example, cuts on the transverse
momentum, rapidity, and isolation of leptons or photons visible
in the final state.  In an ideal world, a flexible hadron-level
Monte Carlo program, accurate to NNLO, would allow the effects
of such cuts to be assessed. However, even fixed-order computations
of generic (``multi-scale'') NNLO observables are not yet
available.

A special case which can be handled in the style of the previous section is the
distribution in rapidity $Y_V$ of a Drell-Yan pair or a Higgs boson.  The
former process can be used to extract parton distribution information from
fixed-target production, or monitor the ``partonic luminosity'' via $W$ and $Z$
production at the LHC~\cite{DPZ}, because at leading order it is proportional
to $q(x_1) \bar{q}(x_2)$ with $x_{1,2} = (M_V/\sqrt{s}) e^{\pm Y_V}$. Compared
with the total cross section approach of ref.~\cite{AnastasiouMelnikov}, a
$\delta$ function needs to be inserted into the phase-space integration, of the
form $\delta(Y-Y_V)$.  The IBP method still works~\cite{ADMP}, reducing the
phase-space integrals to a set of master integrals depending on $z$ and $u =
(x_1/x_2) e^{-2Y_V}$, which can again be determined by integrating differential
equations. The NNLO results have much lengthier expressions than the one-scale
answers.  They involve polylogarithms with arguments which can be irrational
functions of $z$ and $u$, for example
$\Li_2[(u-1-i\sqrt{(4u^2-z(1+u)^2)/z}/(2u)]$. The same stability of the
perturbative series seen for the total $W$ and $Z$ production cross section at
NNLO, holds also bin-by-bin in rapidity~\cite{ADMP}.

For problems with more than two scales, for example $e^+e^-$ event shapes and
generic hadron-collider processes, a flexible, fully numerical approach seems
necessary.  The major bottleneck at present comes in integrating contributions
containing the emission of two extra partons, which have quite complicated
singularities as momenta become soft and/or collinear.  There has been
important recent progress in this direction.  For lack of space, and because
these developments were described in another talk at this workshop, I refer the
reader to that report~\cite{Klasen}.


\section{Twistor spinoffs}
\label{twistorsection}

There has been a great deal of recent interest in novel
methods for evaluating QCD tree and loop amplitudes,
stimulated by Witten's topological string in twistor
space~\cite{WittenTopologicalString}.  In general,
it is possible to find compact representations for
amplitudes for many external particles, and more efficient
techniques for computing the amplitudes, by making
full use of their analytic structure (which is sometimes hidden).
Twistor space~\cite{Penrose}
is a kind of Fourier transform of the usual
momentum-space representation of amplitudes.  Very often,
a Fourier transform can expose simplicity.  Consider the time-dependence
of the electric field $E(t)$ associated with light arriving from the Sun.
It has a pretty random appearance.  However, transforming to
energy variables, $E(t) \to E(\omega) = \int dt e^{i\omega t} E(t)$,
reveals spectral lines, from which the presence of helium in
the Sun could be deduced.

The twistor transform is very well-suited for describing the scattering
of massless particles.  Traditional scattering variables are the
four-momentum vectors $k_i^\mu$ --- which are null vectors in the massless
case, $k_i^2 = 0$ --- and their Lorentz-invariant products,
$s_{ij} = (k_i+k_j)^2 = 2k_i\cdot k_j$.
We can trade the $k_i^\mu$ for spinor variables, the
right- and left-handed, or $+$ and $-$ chirality, solutions
to the Dirac equation, $u_{\pm}(k_i)$.  A shorthand notation
for the two-component (Weyl) versions of these spinors is,
\begin{equation}
(\lambda_i)_\alpha \equiv u_+(k_i),
\qquad
(\tlambda_i)_{\dot\alpha} \equiv u_-(k_i).
\end{equation}
The trade is possible thanks to the form of the positive-energy
projector for massless spinors, $u(k)\bar{u}(k) = \ksl$,
or in two-component notation,
\begin{equation}
k_i^\mu (\sigma_\mu)_{\alpha\dot\alpha}
= (\ksl_i)_{\alpha\dot\alpha}
= (\lambda_i)_\alpha (\tlambda_i)_{\dot\alpha} \,.
\label{kfact}
\end{equation}

Instead of Lorentz-invariant products, the natural variables
for massless scattering are spinor inner-products~\cite{MPReview},
defined by
\begin{equation}
\spa{j}.{l}
= \ve^{\alpha\beta} (\lambda_j)_\alpha (\lambda_l)_\beta
= \bar{u}_-(k_j) u_+(k_l)\,,
\qquad
\spb{j}.{l}
= \ve^{\dot\alpha\dot\beta} (\tlambda_j)_{\dot\alpha} (\tlambda_l)_{\dot\beta}
= \bar{u}_+(k_j) u_-(k_l)\,.
\label{spinorproddef}
\end{equation}
These products are the square roots of the Lorentz products, up to a
phase $\phi$,
\begin{equation}
\spa{j}.{l}
= \sqrt{s_{jl}} e^{i\phi_{jl}} \,,
\qquad
\spb{j}.{l} = \pm \sqrt{s_{jl}} e^{-i\phi_{jl}} \,.
\label{spinorprodtwo}
\end{equation}
The utility of these variables was recognized already in the 1980s.
For example, the Parke-Taylor tree amplitudes~\cite{ParkeTaylor,BGSixMPX}
are for the scattering of two negative-helicity gluons, labelled $j$ and $l$,
and $n-2$ positive-helicity gluons.  They are termed ``maximally
helicity-violating'' (MHV) amplitudes, because tree amplitudes with
fewer (zero or one) negative-helicity gluons vanish.  In terms of
spinorial variables, they take a remarkably simple form for any $n$,
\begin{equation}
A_n^{{\rm MHV}\,,jl}
\equiv
A_n^{\rm tree}(1^+,2^+,\ldots,j^-,\ldots,l^-,\ldots,n^+)
= i { {\spa{j}.{l}}^4 \over \spa1.2\spa2.3\cdots\spa{n}.1 } \,,
\label{PTAmps}
\end{equation}
depending only on the positive-helicity spinors $\lambda_i$,
not the $\tlambda_i$.

\subsection{Twistor space and MHV rules}

\begin{figure}
  \includegraphics[height=.18\textheight]{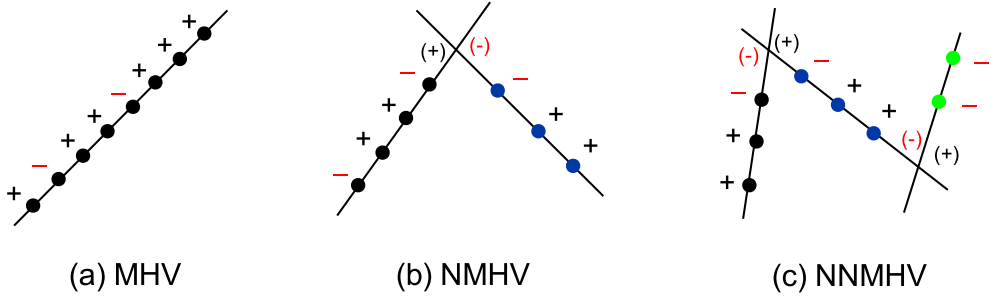}
  \caption{Tree amplitudes are supported on networks of intersecting
 lines in twistor space.}
 \label{mhvtreefigure}
\end{figure}

The twistor transform is a Fourier transform of the $\tlambda_i$,
leaving the $\lambda_i$ alone.  The four coordinates of
twistor space, for each of the $n$ particles, are
$(\lambda_1,\lambda_2,\mu^{\dot1},\mu^{\dot2})$,
where $\mu^{\dot\alpha}$ is defined by
\begin{equation}
\tlambda_{\dot\alpha} = i { \del \over \del \mu^{\dot\alpha} } \,,
\qquad
\mu^{\dot\alpha} = i { \del \over \del \tlambda_{\dot\alpha} } \,.
\label{mudef}
\end{equation}
In order to transform the MHV amplitudes~(\ref{PTAmps}),
following ref.~\cite{WittenTopologicalString}
we first must multiply them by the momentum-conserving $\delta$-function,
which can be written, using \eqn{kfact} as,
\begin{equation}
\delta\Bigl( \sum_i k_i \Bigr)
= \int d^4x
\exp[ i x^{\alpha\dot\alpha}
(\lambda_i)_\alpha (\tlambda_i)_{\dot\alpha} ] \,.
\label{deltafn}
\end{equation}
Then the transformed amplitudes are
\begin{equation}
\tilde{A}_n^{{\rm MHV}\,,jl}(\lambda_i,\mu_i)
 = \int \prod_i d\tlambda_i \exp( i \mu_i \tlambda_i )
  \int d^4x\ A(\lambda_i)
 \exp[ i x \lambda_i \tlambda_i ]
\propto \prod_i \delta( \mu_i + x \lambda_i ) \,.
\label{TwistorMHV}
\end{equation}
The product of all the linear $\delta$ functions means
that the amplitude is supported on a line in
twistor space, as shown in \fig{mhvtreefigure}(a).

\begin{figure}
  \includegraphics[height=.14\textheight]{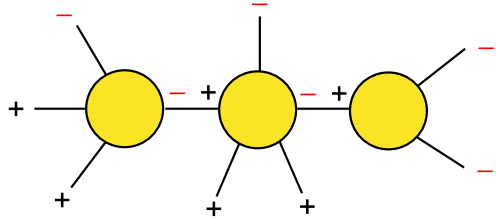}
  \caption{Example of an MHV-rules diagram, corresponding to
  \fig{mhvtreefigure}(c).}
 \label{mhvdiagram}
\end{figure}

Investigation of amplitudes with three and four negative helicities (NMHV and
NNMHV amplitudes) revealed the pattern of intersecting lines in
\fig{mhvtreefigure}(b) and \fig{mhvtreefigure}(c).  It also led to a set of
``MHV rules'' for QCD tree amplitudes, which are simpler than Feynman
rules~\cite{CSW}. Each line in \fig{mhvtreefigure} corresponds to an ``MHV
vertex'', which is a clever off-shell continuation of the MHV
amplitude~(\ref{PTAmps}), labelled with two negative helicities and the rest
positive.  Many Feynman vertices can be lumped effectively into a single MHV
vertex.  The MHV vertices are joined with scalar propagators, that is, factors
of $1/p^2$, so that no messy contractions of Lorentz indices have to be
performed.  An example of an MHV-rules diagram, corresponding to the
twistor-space structure in \fig{mhvtreefigure}(c), is given in
\fig{mhvdiagram}.

The efficiency of the MHV rules for gluonic amplitudes
motivated their quick extension to amplitudes with massless
external fermions~\cite{ExtFerm},
Higgs bosons coupling to gluons via $H\tr(G_{\mu\nu}G^{\mu\nu})$ in
the large $m_t$ limit~\cite{ExtHiggs}, and vector bosons ($\gamma^*,W,Z$),
including DIS multi-jet processes~\cite{ExtVector}.

\begin{figure}
  \includegraphics[height=.17\textheight]{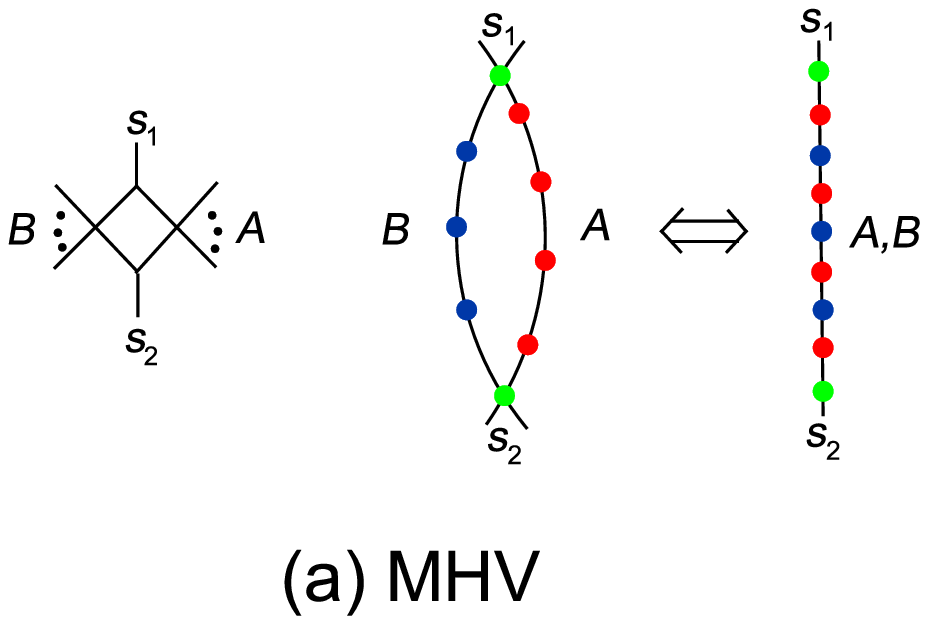}
  \hskip1.5cm
  \includegraphics[height=.17\textheight]{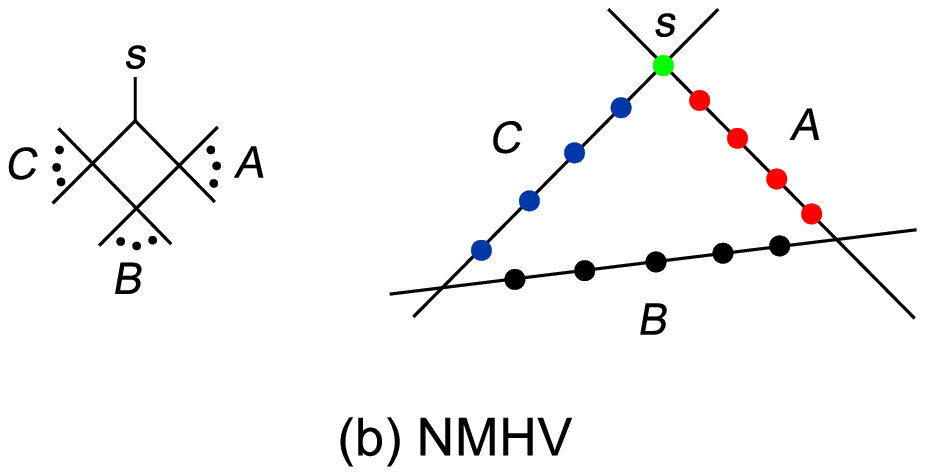}
  \caption{Twistor structure of box integral coefficients for
one-loop amplitudes in ${\cal N} = 4$ supersymmetric Yang-Mills theory.}
 \label{looptwistfigure}
\end{figure}

In a parallel development, the twistor structure of loop amplitudes
was explored and exploited~\cite{TwistorLoops}.
The simplest structure to explain
is for gluonic loop amplitudes in a computational ``toy model''
for QCD, maximally (${\cal N} = 4$) supersymmetric Yang-Mills theory.
One-loop amplitudes in this theory can be written as a linear combination
of scalar box integrals; no triangles or bubbles are required.
The box integrals typically have many legs from the amplitude
clustered into a single vertex of the box.
In the MHV case, the coefficients of the boxes are either zero,
or else equal to
the MHV tree amplitudes $A_n^{{\rm MHV}\,,jl}$~\cite{Neq4MHV},
in which case the vertices all lie on a single line in twistor space,
as shown in \fig{looptwistfigure}(a).
However, this pattern turns out to be a degenerate case, which
is resolved in the NMHV amplitudes.  Here the simplest non-vanishing
box coefficients are those with three clusters
$A,B,C$ and a single massless leg $s$. Their coefficients are
supported on three lines intersecting in a ring, as shown
in~\fig{looptwistfigure}(b)~\cite{Neq4NMHV}.

\subsection{On-shell recursion relations}

Quite recently, another approach to tree amplitudes
has been developed, on-shell recursion relations~\cite{RSV,BCF,BCFW}.
These relations are somewhat more efficient and easier to
generalize than the MHV rules~\cite{RecTrees}, and they have very promising
implications for loop amplitudes as well~\cite{RecLoops}.
The derivation of the relations~\cite{BCFW} is very general,
relying just on Cauchy's theorem for functions of
a single complex variable, and the factorization properties
of amplitudes.  The desired amplitude $A_n = A_n(0)$ can be
embedded into a family of amplitudes $A_n(z)$, labelled by a
complex parameter $z$ characterizing a shift in the momentum
flowing through the amplitude.  For example, the momenta
of the pair of legs 1 and $n$ can be shifted, while respecting
overall momentum conservation and masslessness of the external
legs, according to $k_1 \to {\hat k}_1(z)$, $k_n \to {\hat k}_n(z)$,
where
\begin{equation}
 {\hat k}_1(z) + {\hat k}_n(z) = k_1(z) + k_n(z),
 \qquad
  {\hat k}_1^2(z) = {\hat k}_n^2(z) = 0.
\label{shift1n}
\end{equation}

A momentum shift satisfying \eqn{shift1n} is best described using spinor
variables, as
\begin{equation}
 \lambda_1 \to \lambda_1 + z \lambda_n, \quad
  \tlambda_1 \to \tlambda_1;
\qquad
 \lambda_n \to \lambda_n, \quad
  \tlambda_n \to \tlambda_n - z \tlambda_1.
\label{spinorshift1n}
\end{equation}
This is because a complex massless vector $k^\mu$ has $\det(\ksl)=k^2=0$.  So
the singular $2\times2$ matrix $\ksl$ can be factored into a pair of spinors,
as $(\ksl)_{\alpha\dot\alpha} = \lambda_\alpha \tlambda^{\prime}_{\dot\alpha}$,
as in \eqn{kfact}, where $\tlambda^\prime$ is no longer the conjugate spinor to
$\lambda$.

As $z$ varies over the complex plane, different intermediate states go on
shell, generating poles in $z$. Let $K_{1,l} \equiv k_1+k_2 + \cdots + k_l$.
Then $\hat{K}_{1,l}^2(z) = (K_{1,l} + z \lambda_n \tlambda_1)^2$ vanishes at
\begin{equation}
z = z_l = - K_{1,l}^2  / \langle n^- | \Ksl_{1,l} | 1^-\rangle  \,.
\label{zl}
\end{equation}
So long as $A_n(z) \to 0$ as $z \to \infty$, the contour
integral over a large circle $C$, ${1\over2\pi i}\oint_C dz A_n(z)/z$
vanishes.   The residue at $z=0$, which is the desired amplitude
$A_n(0)$, is the negative of the sum of the residues at $z=z_l$.
Those residues are given by the factorization of the amplitude
into two lower-point amplitudes, evaluated in shifted, on-shell
kinematics.  The resulting recursion relation~\cite{BCF}
includes a sum over $l$, and over the possible intermedate
helicities $h$,
\begin{equation}
A_n(1,2,\ldots,n) =
\sum_{h=\pm} \sum_{l=2}^{n-2}
A_{l+1}(\hat{1},2,\ldots,l,-\hat{K}_{1,l}^{-h})
{i\over K_{1,l}^2}
A_{n-l+1}(\hat{K}_{1,l}^h,l+1,\ldots,n-1,\hat{n}).
\label{OSRR}
\end{equation}

\begin{figure}
  \includegraphics[height=.17\textheight]{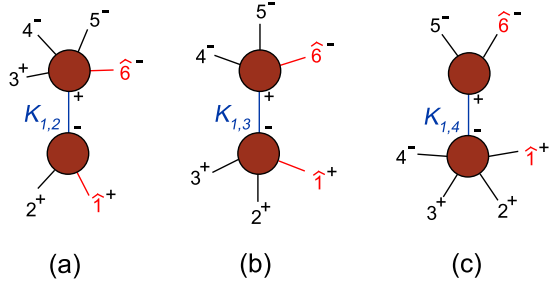}
  \caption{On-shell recursive diagrams for $A_6(1^+,2^+,3^+,4^-,5^-,6^-)$.}
 \label{BCFpppmmmfigure}
\end{figure}

These relations lead quickly to very compact forms for tree amplitudes.  For
example, there are 220 Feynman diagrams for the six-gluon amplitude.  Using
color algebra and symmetries, the information in these diagrams is represented
by the MHV amplitudes~(\ref{PTAmps}), plus two more helicity amplitudes,
$A_6(1^+,2^+,3^+,4^-,5^-,6^-)$ and $A_6(1^+,2^+,3^-,4^+,5^-,6^-)$. Computing
the first of them using the shift~(\ref{spinorshift1n}), yields the set of
diagrams shown in \fig{BCFpppmmmfigure}. Diagram (b) vanishes because
$A_4({-},{+},{+},{+}) = 0$. Diagram (c) is related to diagram (a) by the
symmetry $(1\lr6,2\lr5,3\lr4)$ (plus spinor conjugation). Diagram (a) can be
evaluated in a few steps, using the MHV amplitudes, to give a single-term
expression.  Adding diagram (c) gives,
\begin{eqnarray}
-i A_6(1^+,2^+,3^+,4^-,5^-,6^-) &=&
 { { \langle 6^- |(1+2) | 3^- \rangle }^3
   \over \spa6.1 \spa1.2 \spb3.4 \spb4.5 s_{612}
\langle 2^- |(6+1)|5^-\rangle }
\nonumber \\
&&\hskip0.05cm +  { { \langle 4^- |(5+6) | 1^- \rangle }^3
   \over \spa2.3 \spa3.4 \spb5.6 \spb6.1 s_{561}
\langle 2^- |(6+1)|5^-\rangle }  \,.
\label{pppmmmsimple}
\end{eqnarray}
The combination
$\langle 2^- |(6+1)|5^-\rangle = \spa2.6 \spb6.5 + \spa2.1 \spb1.5$
in the denominator leads to an unphysical singularity in the first
term of \eqn{pppmmmsimple} when $k_6+k_1$ is a linear combination
of $k_2$ and $k_5$, which is cancelled by the second term.
On the other hand, all of the physical factorization behavior is
made manifest, in contrast to Feynman-diagram based
representations~\cite{MPReview}.
The amplitude $A_6(1^+,2^+,3^-,4^+,5^-,6^-)$ has three independent
recursive diagrams. Thus the six-gluon calculation is reduced
from 220 Feynman diagrams to 4 much simpler ones.

The generality of this approach has led to its rapid application to different
tree-level processes, with fermions and massive particles~\cite{RecTrees}, and
to one loop~\cite{RecLoops}. Essentially, amplitudes are being built up
directly from their analytic properties. Considering that the heyday of
$S$-matrix analyticity ended with the rise of a gauge theory for the strong
interactions, QCD, these recent computational advances may herald the final
revenge of the analytic $S$-matrix.


\begin{theacknowledgments}
I am grateful to the organizers of DIS 2005 for the invitation to
present this talk, and for arranging such a stimulating meeting.
I thank Zvi Bern, Vittorio Del Duca, Michael Klasen and David Kosower
for helpful discussions.
\end{theacknowledgments}

\end{document}